\documentclass[11pt]{article}

\usepackage{amssymb} 
\usepackage{amscd}
\usepackage{xypic}
\usepackage[matrix,arrow,curve]{xy}

\def \be {\begin{equation}}
\def \ee {\end{equation}}
\def \bea {\begin{eqnarray}}
\def \eea {\end{eqnarray}}
\def \ba {\begin{array}}
\def \ea {\end{array}}

\def \zz {{\Bbb Z}}

\def \aa {{\cal A}}

\def \oo {{\cal O}}

\def \hh {{\cal H}}

\def \CY {Calabi-Yau}
\def \Ka {K\"ahler}

\def \Ai {$A_\infty$}
\newcommand {\mat}[1]{\mathop{\rm #1}\nolimits}

\newcommand{\Hom}{\mathop{\rm Hom}\nolimits}
\newcommand{\Ext}{\mathop{\rm Ext}\nolimits}

\begin{document}

\begin{flushright}
SISSA 59/2001/FM\\
hep-th/0107195\\
\end{flushright}

\begin{center}
{\Large\bf
A-infinity structure and superpotentials
\\}
\vspace{.8cm}
{\small Alessandro Tomasiello}\\
\vspace{.6cm}
{\small
International School for Advanced Studies (SISSA),\\
via Beirut 2-4, 34014 Trieste, Italy\\
\vspace{.4cm}
{\tt tomasiel@sissa.it}}

\vspace{.8cm}

\begin{abstract}
\noindent
\Ai\ algebras and categories are known to be the algebraic structures behind
open string field theories. In this note we comment on the relevance of 
the homology construction of \Ai\ categories to superpotentials.
\end{abstract}

\end{center}

\vspace{1cm}

Branes in \CY\ manifolds
are an important arena for uncovering nonperturbative
features of string theory, with interesting mathematical phenomena as
by-products or important ingredients. Such a sentence can be partially 
justified noting the remarkable fact that mathematicians, trying to uncover
the core of the mirror symmetry phenomena, came to formulate a conjecture
in terms of derived categories first \cite{K}, and later of \Ai\ ones \cite{P}.
The physical reader of such conjectures comes immediately to think that
the latter
have to do in some way with string field theory, in which \Ai\ was 
recognized to emerge some time ago \cite{Gaberdiel:1997ia}. 
This note is a first try to make this connection
more precise, applying it to the problem of superpotentials in \CY\ 
compactifications \cite{Brunner:2000jq,supot}.

\vspace{.3cm}
\noindent
{\it \Ai\ and mirror symmetry}
\vspace{.2cm}

Everything starts from the study of twisted topological models 
and their boundary conditions \cite{Witten:1992fb}. 
Remarkably, A and B branes make an {\sl
ante litteram} appearance there. This is one of the main reasons leading to 
a conjecture \cite{K} meant to explain the ``mathematical mysteries'' of mirror
symmetry. This {\it homological mirror symmetry} conjecture states 
the equivalence of the derived category of
coherent sheaves on a \CY\ $Y$ and of the derived category of the Fukaya 
category of the mirror $\tilde Y$. This relates respectively B branes on
$Y$ with A branes on $\tilde Y$. 

But, Fukaya category is an \Ai\ category. Since \Ai\ structure is
important in string field theory, to a physicist this 
should already suggest the existence of a refined conjecture making reference
directly to the Fukaya category itself, without 
having to associate a derived category to it. Remarkably, even without this
string field theory suggestion, mathematicians came to the same conclusion,
formulating (and checking in one example) exactly this refined conjecture 
\cite{P}, 
building on previous general work on how to define an \Ai\ structure 
from differential graded algebras \cite{M}. 

What a physicist can do now is to try to use this {\sl explicit} \Ai\ structure
 to
do computations in string theory. What we know is that a string field theory
action $S$ obeying Batalin-Vilkovisky master equation $\{ S, S\}=0$
can be written from an \Ai\ structure and a bilinear form. Reviewing 
this requires a crash course on \Ai\ structures.

\vspace{.3cm}
\noindent
{\it \Ai\ and string field theory}
\vspace{.2cm}

Let us start from \Ai\ algebras \cite{St}. These are generalizations of
differential associative algebras, in which the product is not required
to be associative. A clever way to introduce them is the so-called bar 
construction \cite{St,ke}, whose relevance in string theory would be 
interesting to 
elucidate. We will instead be more down to earth and define \Ai\ algebra
as a $\zz$-graded vector space $\aa$, endowed with linear maps 
\[m_k: \underbrace{\aa\otimes\ldots\otimes \aa}_{k \mat{times}}\to \aa \]  
of grade $2-k$, satisfying an infinite set of conditions, among which we 
display only the first ones:
\bea
\nonumber &&m_1^2= 0\ ; \qquad 
m_1\big(m_2(\phi_1,\phi_2)\big)= m_2\big(m_1(\phi_1), \phi_2\big)+
(-)^{|\phi_1|}m_2\big(\phi_1,m_1(\phi_2)\big) ; \\
&&m_2\big(\phi_1,m_2(\phi_2,\phi_3)\big)
-m_2\big(m_2(\phi_1,\phi_2),\phi_3\big)=
m_1\big(m_3(\phi_1,\phi_2,\phi_3)\big)+\label{Ainf}\\
\nonumber &&
m_3\big(m_1(\phi_1),\phi_2,\phi_3\big)+ 
(-)^{|\phi_1|}m_3\big(\phi_1,m_1(\phi_2),\phi_3 \big)+
(-)^{|\phi_1|+|\phi_2|}m_3\big(\phi_1,\phi_2,m_3(\phi_3)\big)\ ,
\eea
with $\phi_i\in \aa$, and $|\phi_i|$ their grades. 
Second equation shows $m_1$ is a differential; $m_2$ can be thought of as
a multiplication which is not associative but almost so, as measured by the
presence of the terms with $m_3$ in (\ref{Ainf}).

If one has such a structure on the Hilbert space of a string theory (and a
symplectic bilinear form $\langle\,,\rangle$), one can 
define the promised string field theory action as
\be
 S= \frac12\langle\Phi, Q \Phi \rangle + 
\frac13 \langle \Phi, b_2(\Phi,\Phi)\rangle + \frac14 \langle \Phi,
b_3(\Phi,\Phi,\Phi)\rangle + \ldots
\label{S}
\ee
where actually the $b_k$ are not the $m_k$ but close relatives 
\cite{ke,Gaberdiel:1997ia}, and $\Phi$ is the string field. One can
recognize in this formula the differential $m_1$ as being the BRST operator 
$Q$, and $m_2$ as being the string field theory product.

A similar work, generalizing associative structure to \Ai\ one, 
can be done with linear categories. The idea of linear
categories is just to relax the usual conditions on product structures in
algebra, that require the product to be there {\sl for any} couple of elements.
The way to do this is to introduce first a class of useful labels, called
{\it objects}.
Then, one introduces for any pair of objects $a, b$, a vector space
of {\it morphisms}
$\Hom(a,b)$. These are the generalization of the elements of the algebra:
there is a product $\circ$ such that a $\phi \in \Hom(a,b)$ and a
$\phi'\in\Hom(a',b')$ can be multiplied only if $b=a'$: one can imagine the
morphisms as arrows, and say that the product is defined only when the head of
the first arrow coincides with the tail of the second one. With this important
proviso, which is the whole difference between a category and an algebra, all
the rest of the definition remains the same: namely, in all the $\Hom(a,b)$ it
is required to have an unity element; there should be associativity $(\phi
\circ \phi') \circ \phi''= \phi
\circ (\phi' \circ \phi'')$ whenever the products are defined; and $\circ$
should be bilinear in the two entries (distributive law).

Now, as for algebras, one can define an \Ai\ category by replacing 
associative multiplication with maps 
\[ m_k : \Hom (B_1,B_2) \otimes \Hom(B_2,B_3)\otimes\ldots \otimes 
\Hom(B_{k-1},B_k) \to \Hom(B_1,B_k)\]
obeying conditions (\ref{Ainf}) whenever the multiplications are defined.
This is relevant for physics again thanks to the applications to string field
theory \cite{laz}. Indeed, branes form 
categories: This means that the objects of the
category are {\sl branes}, and the morphisms between them are the states in
the Hilbert space of open strings connecting them. In general this category
will be an \Ai\ one;
then the analogue of (\ref{S}) can be defined in this case \cite{laz}, 
using the more refined
\Ai\ category structure instead of the \Ai\ algebra structure. 
It is
more refined in the sense that, given a category, one can always do a trick
and consider the algebra obtained ``collapsing all objects to one''; namely,
defining a product also for arrows whose head and tail do not coincide, as a
zero product. One can do so, but the category picture displays more structure.

\vspace{.3cm}
\noindent
{\it Homology and superpotentials}
\vspace{.2cm}

We have said branes form a category. If one considers off-shell
states, the category will sometimes be associative (as in the case of 
topological models), but in general will be an \Ai\ one. For any \Ai\ algebra 
$\aa$, we can define its {\it homology} $H^*\aa$, whose objects are the same 
as those
of $\aa$ and whose morphisms are the cohomology of $\aa$ with respect to the 
differential $m_1$. This is again an \Ai\ algebra \cite{ke}; if $\aa$ is a
differential graded algebra this can be explicitly shown \cite{ka}, 
as we are going to review in a moment. Again, all these things can be 
generalized to categories.

Physically, $H^* \aa$ is the Hilbert space of physical states, and this is 
where things start being interesting. Let us consider a definite case: 
B branes in a \CY\ threefold $Y$. 
We know that the problems of classifying such states
at a general point in moduli space decouples in a holomorphic problem, coming
from an F-flatness in 4d terminology, and a more difficult 
stability problem, coming from the D-term; they depend on holomorphic and \Ka\
moduli respectively \cite{Brunner:2000jq}. 
To talk about the holomorphic side without having to
worry about stability, a standard trick is to consider topologically B-twisted
model. In this case, string field theory action has been shown 
\cite{Witten:1992fb} to be 
associative, and more precisely to have the form of holomorphic CS action
\be
 S= \int \Omega\wedge\left(\frac12 A \wedge \bar\partial A + 
\frac13 A\wedge A \wedge A\right)\label{holCS}
\ee
for space filling branes, with $A$ being $(0,1)$ connection. In the
non-space-filling cases, one takes dimensional reduction, as for instance in
\cite{supot}.

Now, the crucial idea is this. The string field theory (\ref{S})
action is determined by the algebraic structure (the $m_k$). 
This action gives in principle
a means to compute all of the correlation functions. What is the meaning of 
the algebraic structure on the category of physical states, then? They will be
correlation functions of the physical states. Since this is the same as
computing superpotentials (see for example \cite{Brunner:2000jq} for details),
we can then interpret these as giving term by term the complete 
superpotential of the theory.

Let us come back to our example, and compute this induced \Ai\ structure on the
physical states. For simplicity we start taking two space-filling branes $B_0$
and $B_1$. 
The power of this method is however that it can be extended to coherent
sheaves and more
general elements of the derived category; in perspective, 
this can be useful away from the
large volume limit, where one expects that ``strange'' elements of the 
derived category will become stable. 
The starting Hilbert space is simply \cite{Witten:1992fb} 
$\Omega^{(0,p)}(B_1^*\otimes B_2)$ with
its structure of differential graded algebra, the differential being
$\bar\partial_{B_1^*\otimes B_2}\equiv\bar\partial$ and the product being the
$\wedge$ product between forms. The physical states are
then the cohomology of the $\bar\partial$ operator. 
 Note that the holomorphic CS action is
simply the application of (\ref{S}) to this algebra, and this tells us that
in this case the symplectic form on string field algebra is
$\int\Omega(\cdot\wedge\cdot)$.

The \Ai\ structure on this algebra is now given as follows: Consider
the operator $\hh \equiv 1 - \Delta_{\bar\partial}G - G\Delta_{\bar\partial}$.
This is the projector on harmonic forms \cite{GH,M,P}. We recall that 
harmonic forms are
a slice of the cohomology, in the sense that there is exactly one harmonic
form in any cohomology class. This is familiar: they satisfy 
$\bar\partial \omega=0=\bar\partial^\dagger\omega$; first condition is
their being closed, second one ``fixes the gauge'' of the invariance $\omega \cong
\omega + \bar\partial \alpha$. The vector space on which we will impose
the \Ai\ structure is thus the cohomology of the complex, or equivalently the
subcomplex of harmonic forms, $\Ext^p(B_1,B_2)$. Now, the products. The 
differential $m_1$ is taken to be zero (it is $\bar\partial$ restricted to 
harmonic forms). The product $m_2$ is simply the harmonic part of the wedge 
product:
\[ m_2(A_1, A_2) = \hh(A_1 \wedge A_2)\ .\]
Higher products are less obvious, but are the interesting part. They can 
be written iteratively 
in terms of the operator $p\equiv \bar\partial^\dagger G$ (where $G$ is 
Green function for the the laplacian $\Delta_{\bar\partial}$ as \cite{M}
\bea\nonumber m_k(A_1,\ldots,A_k)&\equiv&\\
\nonumber&\hh&\Big\{ 
(-)^{k-1}[p\,m_{k-1}(A_1,\ldots,A_{k-1})]\wedge A_k
- (-)^{k\, a_1}A_1 \wedge[p\,m_{k-1}(A_1,\ldots,A_{k-1})]\\
\nonumber &&- \,\sum_{i+j=k+1\atop i,j\geq2}
(-1)^{i+(j-1)(a_1+\ldots+
a_i)} [p\,m_i(A_1, \ldots, A_i)] \wedge [p\,m_j(A_{j+1},\ldots,A_k)]\Big\}
\eea
where $a_i$ is the degree of $A_i$. Having such products in our hands, 
the superpotentials will be now, exactly in analogy with the string field 
action (\ref{holCS}), and based on what we have said before, 
\be
 W= \int\Omega\wedge\mat{tr}\left\{
A\wedge(\frac13A\wedge A+\frac14m_3(A,A,A)+
\frac15m_4(A,A,A) +\ldots )\right\} ,\label{W}
\ee
where $A$ are now {\sl harmonic} $(0,1)$ forms. Actually, more precisely, 
these $A$ have to be understood as elements of $\Ext^1(B_1,B_2)$, though we
will keep the notation for forms. In 
particular, for lower-dimensional branes some of these $A$ have to be
understood actually as transverse scalars $X$, as we will see more precisely
later. Note also that trace in (\ref{W}) is simply circular matching of
gauge indices of the various ``bifundamentals'' corresponding to strings
stretched between $B_1$ and $B_2$, between $B_2$ and $B_3$,\ldots, between
$B_k$ and $B_1$.

Let us see how do really these higher products look like by working out $m_3$.
The corresponding piece of the superpotential is simply of the form (in a
little schematic way again, for the time being)
\be 
\int\Omega\wedge \mat{tr}\left\{A\wedge A\wedge\bar\partial^\dagger (G\,A
\wedge A )\right\}\ ;
\label{fey}
\ee
we have written the operator $p$ explicitly to underline the structure of this
piece. $G$ can be understood roughly as $\Delta^{-1}$, and so 
the operator $\bar\partial^\dagger G $ is formally a kind of 
$(\bar\partial)^{-1}$,
that is, the propagator corresponding to the first-order kinetic term
$\bar\partial$ (or, if one prefers, in string field theory terms this is
$b_0/L_0$). The term (\ref{fey}) can thus be thought of as the Feynman 
diagram 
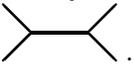
\setlength{\unitlength}{0.25mm}
\begin{picture}(60,30)
\thicklines
\put(0,0){\line(1,1){15}}
\put(0,30){\line(1,-1){15}}
\put(15,15){\line(1,0){30}}
\put(45,15){\line(1,1){15}}
\put(45,15){\line(1,-1){15}}
\end{picture}\ .

One can see a similar Feynman interpretation for all the terms in this
superpotential; 
this is more or less the reinterpretation of the \Ai\ construction
\cite{M} in terms of trees \cite{Kontsevich:2000yf}, and is the better 
justification for our claims: in a sense the \Ai\ structure has the role of
resumming all the graphs (something similar happens in \cite{Nakatsu:2001da}).
It also suggests a more general 
relevance of \Ai\ structure in effective field theories.

In these expressions, again these $A$ should be thought of as being a symbol
for several things. We will try now to be more precise. First of all, as we
mentioned, for lower-dimensional branes these $A$ should be understood as $A$'s
or transverse scalars $X$'s. Both of these are elements of
$Ext^1(B,B)$. Mathematically, one can see this in the following way. Consider
for simplicity a bundle on a divisor $D$ which is restriction of a bundle $E$ 
on the ambient manifold $Y$. Then from the usual exact sequence
\[ 0 \to E \otimes\oo(-D)\to E \to E_{|_D} \to 0 \]
and considering the exact sequence of $\Ext^i(\,\cdot\,, E_{|_D})$ we get 
\[ \ldots \to H^0(D, End(E))\otimes N_{D,Y} \to \Ext^1(E_{|_D},E_{|_D})\to
H^1(D,End(E))\to\ldots\ ;\]
in this sense, $\Ext^1$ is made of these two pieces, transverse scalars and
connections, as one already knows from reduction common sense. The fact that
we consider here $\Ext^1$ is justified by the fact that these are the internal
parts of the chiral multiplets, whose superpotential we are indeed computing.

Apart from this, we should note that in easy situations one can get 
from this somewhat trivial results; the various
terms in each $m_k$ may cancel (we have schematically displayed only one in
(\ref{fey})). In more general situations, however, these will give probably
interesting results about obstructions of moduli spaces of curves. Just to
have an idea, a typical term arising by reduction on e.g. a curve will be
\[ \int\Omega_{ijz}X^iA_{\bar z}\,\bar\partial^\dagger( G X^j A) dz\wedge
d\bar z\ ,\]
where this time we have denoted by $A$ really connections, being instead
transverse scalars explicitly denoted by $X^i$. 
Let us also note again that this formalism can be applied more generally to
$A\in\Ext^1(B_i,B_j)$ with $B_i$ general objects in the derived category 
\cite{P}.

An important direction of development is towards inclusion of closed string
theory. The closed string field theory for the B model, in analogy for the
open string sector, is known \cite{Bershadsky:1994cx} as 
Kodaira-Spencer theory, because it reproduces the homonymous deformation 
theory. Coupling between the two encodes deformations of open string field 
theory \cite{Hofman:2001ce}. The idea should be 
to generalize the homology construction to
the BV algebra of open-closed string theory; namely, to find another such
structure on the homology of the open-closed algebraic structure, generalizing
the Kadeishvili theorem we have cited here for \Ai\ algebras (or categories).
This will probably have again a Feynman diagram interpretation, and would
allow us to interpret the result as open-closed superpotentials. Such terms
are very interesting because describe the behaviour of embedded families 
of subvarieties varying complex structure moduli \cite{Brunner:2000jq}.

Though we have dealt so far with B branes, analogous reasonings can 
be done with three-dimensional lagrangian submanifolds; in that case string
field theory action is this time usual Chern-Simons theory, with instanton
corrections. $\bar\partial$ is
now replaced by $d$, and physical states are now in $H^i(M, End(E))$, where $M$
is the lagrangian submanifold and $E$ is the flat bundle over it. We can
repeat now the discussion above; finally we should find contact
with Floer homology and \Ai\ structure of Fukaya category.

A last point is the following. An important property of
\Ai\ algebras is their connection with extended moduli spaces: given an
associative algebra $A$, if its second Hochschild $HH^2(A,A)$ parameterizes 
associative deformations, all of them, $HH^*(A,A)$, parameterize \Ai\
deformations. Thus, extended moduli spaces, suggested
in mathematics again from mirror symmetry considerations \cite{K}, 
should indeed be relevant for string theory: they should parameterize 
deformations of the string theory action.
For instance, one can deform the very string field theory algebra of the
topological models from the differential graded category structure 
$\Ext^*(B_i,B_j)$ to an \Ai\ structure, yielding a new action (through the
general formula (\ref{S}) adapted to brane categories in the spirit of
\cite{laz}) which is still interpretable as a string theory. 

{\bf Acknowledgments.} I would like to thank M.~R.~Douglas, T.~Jayaraman, 
S.~Merkulov, R.~Minasian, V.~Schomerus, S.~Theisen for correspondence and 
discussions.

{\bf Note added.} While this note was in preparation, a paper appeared
\cite{scoop!} in which the same results are derived.


\begin{thebibliography}{99}

\bibitem{K} M.~Kontsevich, {\it Homological algebra of mirror symmetry},
Proceedings of ICM (Z\"urich, 1994), 120--139. Birkh\"auser, Basel, 1995, 
[alg-geom/9411018].

\bibitem{P} A.~Polishchuk, {\it Homological mirror symmetry with higher
    products}, math.AG/9901025.

\bibitem{Gaberdiel:1997ia}
M.~R.~Gaberdiel and B.~Zwiebach,
``Tensor constructions of open string theories I: Foundations,''
Nucl.\ Phys.\ B {\bf 505}, 569 (1997)
[hep-th/9705038].

\bibitem{Brunner:2000jq}
I.~Brunner, M.~R.~Douglas, A.~E.~Lawrence and C.~Romelsberger,
``D-branes on the quintic,''
JHEP {\bf 0008} (2000) 015
[hep-th/9906200].

\bibitem{supot}
S.~Kachru, S.~Katz, A.~E.~Lawrence and J.~McGreevy,
``Open string instantons and superpotentials,''
Phys.\ Rev.\ D {\bf 62}, 026001 (2000)
[hep-th/9912151];

S.~Kachru, S.~Katz, A.~E.~Lawrence and J.~McGreevy,
``Mirror symmetry for open strings,''
Phys.\ Rev.\ D {\bf 62}, 126005 (2000)
[hep-th/0006047];

M.~Aganagic and C.~Vafa,
``Mirror symmetry, D-branes and counting holomorphic discs,''
hep-th/0012041;

I.~Brunner and V.~Schomerus,
``On superpotentials for D-branes in Gepner models,''
JHEP {\bf 0010} (2000) 016
[hep-th/0008194].

\bibitem{St}
J.D. Stasheff, {\it On the homotopy associativity of
$H$-spaces, I.}, Trans. Amer. Math. Soc. {\bf 108}, 275 (1963);
{\it On the homotopy associativity of $H$-spaces, II.},
Trans. Amer. Math. Soc. {\bf 108}, 293 (1963).

\bibitem{Witten:1992fb}
E.~Witten,
``Chern-Simons gauge theory as a string theory,''
hep-th/9207094.

\bibitem{M} S.~Merkulov, {\it Strong homotopy algebras of a K\"ahler
manifold}, math.AG/9809172.


\bibitem{ke} B.~Keller, {\it Introduction to A-infinity algebras and modules}, 
math.RA/9910179.

\bibitem{laz} 
C.~I.~Lazaroiu,
``Unitarity, D-brane dynamics and D-brane categories,''
hep-th/0102183;

C.~I.~Lazaroiu,
``Generalized complexes and string field theory,''
JHEP {\bf 0106}, 052 (2001)
[hep-th/0102122].

\bibitem{ka}
T.~V.~Kadeishvili, {\it The algebraic structure in the homology of an
  $A(\infty)$ algebra}, in Russian, Soobshch. Akad. Nauk Gruzin. SSR {\bf 108}
(1982) 249-252; theorem cited in \cite{ke}.

\bibitem{GH}
Ph.~Griffiths and J.~Harris,
{\it Principles of Algebraic Geometry}, New York, 1978.

\bibitem{Kontsevich:2000yf}
M.~Kontsevich and Y.~Soibelman,
``Homological mirror symmetry and torus fibrations,''
math.sg/0011041.

\bibitem{Nakatsu:2001da}
T.~Nakatsu,
``Classical open-string field theory: A(infinity)-algebra,  renormalization 
group and boundary states,''
hep-th/0105272.

\bibitem{Bershadsky:1994cx}
M.~Bershadsky, S.~Cecotti, H.~Ooguri and C.~Vafa,
``Kodaira-Spencer theory of gravity and exact results for quantum string 
amplitudes,''
Commun.\ Math.\ Phys.\  {\bf 165} (1994) 311
[hep-th/9309140].

\bibitem{Hofman:2001ce}
C.~Hofman and W.~Ma,
``Deformations of topological open strings,''
JHEP {\bf 0101} (2001) 035
[hep-th/0006120].

\bibitem{scoop!}
C.~I.~Lazaroiu,
``String field theory and brane superpotentials,''
hep-th/0107162.

\end{thebibliography}
\end{document}